\documentclass[aps,pra,twocolumn,reprint]{revtex4-1}

\usepackage{amsfonts,amsmath,amssymb}
\usepackage{txfonts}
\usepackage{times}
\usepackage{graphicx}
\usepackage{url}
\usepackage{color}

\usepackage[colorlinks={true},pdftex]{hyperref}  
\hypersetup{citecolor={blue}, filecolor={blue}, linkcolor={blue}, urlcolor={blue},
  colorlinks=true, pdftitle={Enhancement of optimal control yields using quantum controllers}}

\newtheorem{theorem}{Theorem}
\newtheorem{corollary}{Corollary}

\begin{document}
\title{Limits of optimal control yields achievable with quantum controllers}

\author{Re-Bing Wu}\email{rbwu@tsinghua.edu.cn}
\affiliation{Department of Automation, Tsinghua University \& Center for Quantum Information Science and Technology, TNList, Beijing, 100084, China}
\author{Constantin Brif}\email{cnbrif@sandia.gov}
\affiliation{Department of Scalable \& Secure Systems Research, Sandia National Laboratories, Livermore, CA 94550, USA}
\author{Matthew R. James}\email{Matthew.James@anu.edu.au}
\affiliation{ARC Centre for Quantum Computation and Communication Technology, Research School of Engineering, Australian National University, Canberra, ACT 0200, Australia}
\author{Herschel Rabitz}\email{hrabitz@princeton.edu}
\affiliation{Department of Chemistry, Princeton University, Princeton, NJ 08544, USA}
\date{\today}

\begin{abstract}
In quantum optimal control theory, kinematic bounds are the minimum and maximum values of the control objective achievable for any physically realizable system dynamics. For a given initial state of the system, these bounds depend on the nature and state of the controller. We consider a general situation where the controlled quantum system is coupled to both an external classical field (referred to as a classical controller) and an auxiliary quantum system (referred to as a quantum controller). In this general situation, the kinematic bound is between the classical kinematic bound (CKB), corresponding to the case when only the classical controller is available, and the quantum kinematic bound (QKB), corresponding to the ultimate physical limit of the objective's value. Specifically, when the control objective is the expectation value of a quantum observable (a Hermitian operator on the system's Hilbert space), the QKBs are the minimum and maximum eigenvalues of this operator. We present, both qualitatively and quantitatively, the necessary and sufficient conditions for surpassing the CKB and reaching the QKB, through the use of a quantum controller. The general conditions are illustrated by examples in which the system and controller are initially in thermal states. The obtained results provide a basis for the design of quantum controllers capable of maximizing the control yield and reaching the ultimate physical limit.
\end{abstract}
\maketitle

\section{Introduction}

The demand for and interest in quantum control continue to increase with growing successes in manipulating the quantum dynamics of atomic, molecular, optical, and solid-state systems~\cite{Rabitz.Science.288.824.2000, Brif.NJP.12.075008.2010, Brif.ACP.148.1.2012, Dong.Petersen.IET-CTA.4.2651.2010}. Traditionally, the control is implemented by applying a classical electromagnetic field (referred to as a \emph{classical controller}), for example, a laser field applied to a molecular system or a magnetic field applied to a spin ensemble. Recent studies show that control capabilities can be enhanced by coupling the system of interest (referred to as the \emph{quantum plant}) to an auxiliary quantum system (referred to as a \emph{quantum controller}). The use of a suitable quantum controller makes it possible to achieve results beyond those reachable solely by classical fields in applications involving the manipulation of the system's entropy, for example, quantum heat engines~\cite{Allahverdyan.EPL.67.565.2004, Quan.PRE.76.031105.2007, Stefanatos.PRE.90.012119.2014, Kosloff.Levy.ARPC.65.365.2014}, cooling of quantum systems~\cite{Palao.PRE.64.056130.2001, Martin.PRB.69.125339.2004, You.PRL.100.047001.2008, Hamerly.PRL.109.173602.2012}, and quantum error correction with ancillary qubits~\cite{Shor.PRA.52.R2493.1995, *Steane.PRL.77.793.1996, *Bennett.PRA.54.3824.1996, *Gottesman.arXiv.0904.2557.2009}. Such processes can be realized by either direct coupling~\cite{Lloyd.PRA.62.022108.2000, Nelson.PRL.85.3045.2000, Romano.PRA.73.022323.2006, *DAlessandro.IEEE-TAC.57.2009.2009, *Nie.QIC.10.87.2010, *DAlessandro.SCL.62.188.2013, Ticozzi.Viola.SciRep.4.5192.2014} or indirect field-mediated interactions~\cite{Wiseman.Milburn.PRA.49.4110.1994, Yanagisawa.Kimura.IEEE-TAC.48.2107.2003, *Yanagisawa.Kimura.IEEE-TAC.48.2121.2003, James.IEEE-TAC.53.1787.2008, Zhang.James.IEEE-TAC.56.1535.2011, Zhang.James.ChinSciBull.57.2200.2012, Zhang.IEEE-TAC.57.1997.2012, Mabuchi.PRA.78.032323.2008, Iida.IEEE-TAC.57.2045.2012, *Crisafulli.OE.21.18371.2013, Kerckhoff.PRL.109.153602.2012, *Kerckhoff.PRX.3.021013.2013} between the system and a quantum controller, creating \emph{coherent quantum feedback} loops \cite{Zhang.James.ChinSciBull.57.2200.2012, Gough.PTRSA.370.5241.2012} that can mediate the exchange of entropy. A related approach, known as \emph{autonomous feedback}, employs engineered coupling of the controlled quantum system to a dissipative reservoir~\cite{Barreiro.Blatt.Nature.470.486.2011, *Krauter.PRL.107.080503.2011, *Murch.PRL.109.183602.2012, *Geerlings.PRL.110.120501.2013, *Leghtas.PRA.88.023849.2013, *Shankar.Nature.504.419.2013, *Lin.Wineland.Nature.504.415.2013}.

Typical objectives optimized in quantum control applications are the transition probability between the initial and final state and, more generally, the expectation value of an observable (a Hermitian operator on the system's Hilbert space)~\cite{Brif.ACP.148.1.2012}. The value of such an objective achieved through a controlled evolution is commonly referred to as the \emph{control yield}. It is convenient to envision the optimal control design as an excursion over the control landscape defined by the functional dependence of the yield on the control variables. For a classical controller, the search for a globally optimal solution can be very efficient, due to fact that the landscape is free from local optima~\cite{Rabitz.Science.303.1998.2004, *Ho.Rabitz.JPPA.180.226.2006, Rabitz.JCP.124.204107.2006, *Hsieh.JCP.130.104109.2009, Wu.JPA.41.015006.2008} upon satisfaction of reasonable physical assumptions~\cite{Wu.PRA.83.062306.2011, Rabitz.PRL.108.198901.2012, Wu.PRA.86.013405.2012, Riviello.PRA.90.013404.2014, Moore.Rabitz.JCP.137.134113.2012, *Moore.JCP.139.144201.2013}. The vast empirical evidence supporting this result is provided by a large body of successful optimal control experiments~\cite{Brif.NJP.12.075008.2010, Brixner.Gerber.CPC.4.418.2003, *Nuernberger.PCCP.9.2470.2007, Wollenhaupt.Baumert.FaradayDiscuss.153.9.2011} and extensive numerical simulations~\cite{Brif.ACP.148.1.2012, Moore.JCP.128.154117.2008, *Moore.Chakrabarti.PRA.83.012326.2011, *Moore.Rabitz.PRA.84.012109.2011}. Control landscapes for systems with a quantum controller were studied in~\cite{Wu.JMP.49.022108.2008, Pechen.Brif.PRA.82.030101.2010, Wu.Rabitz.JPA.45.485303.2012}, showing that the trap-free property still holds upon satisfaction of similar assumptions over the composite system (which consists of the plant and the controller).

In addition to the topological landscape features such as the presence or absence of local optima, a practically important property is the values of the control objective at the landscape's global maximum and minimum. These values are the \emph{kinematic bounds} on the control yield, which are dynamically reachable provided that the system (or the composite system in the case of a quantum controller) is controllable \cite{DAlessandro.2007.book, Brif.NJP.12.075008.2010, Brif.ACP.148.1.2012}. Since most standard quantum optimal control problems can be formulated as the maximization of the control yield, we are mainly concerned with the upper kinematic bound, but any result obtained for the upper bound can be easily reformulated for the lower one as well. Kinematic bounds have been studied for control of closed~\cite{Girardeau.PRA.55.R1565.1997, *Girardeau.PRA.58.2684.1998, *Schirmer.Leahy.PRA.63.025403.2001} and open~\cite{Khaneja.PNAS.100.13162.2003, Wu.Rabitz.JPA.45.485303.2012} quantum systems. In the case of a closed quantum system coupled to a classical controller, the maximum and minimum yield values are referred to as the \emph{classical kinematic bounds} (CKBs). Since any unitary dynamics cannot change the system's entropy, the CKB is limited and can be surpassed by the addition of a quantum controller that absorbs entropy from the system. For example, it was shown~\cite{Wu.Rabitz.JPA.45.485303.2012} that the kinematic bounds on the fidelity of quantum operations can be improved when the quantum controller is initially in a low-entropy state.

In this paper, we investigate the kinematic bounds on the control yield given by the expectation value of a quantum observable. Relevant applications include, for example, the maximum work that can be exerted by a quantum heat engine~\cite{Allahverdyan.PRE.81.051129.2010} and the maximum degree of purification of the system's state~\cite{Ticozzi.Viola.SciRep.4.5192.2014}. With the use of a suitable quantum controller, the control yield can, in principle, reach the the ultimate physical limit, i.e., the maximum (or minimum) eigenvalue of the target observable, which we refer to as the \emph{quantum kinematic bound} (QKB). Under general circumstances, when the system is coupled to both a classical controller and a quantum controller, the kinematic bound lies between the CKB and QKB. We will present the necessary and sufficient condition for the quantum controller to surpass the CKB, as well as that to attain the QKB.

The paper is organized as follows. In Sec.~\ref{sec:kinematic analysis}, the problem of kinematic bounds on the control yield is formulated for a quantum system coupled to a quantum controller, and topological properties of the associated control landscape are investigated. In Sec.~\ref{sec:qualitative-analysis}, the kinematic bounds are analyzed in the framework of quantum control landscape theory~\cite{Rabitz.Science.303.1998.2004, *Ho.Rabitz.JPPA.180.226.2006, Brif.ACP.148.1.2012}, leading to the derivation of the necessary and sufficient conditions for surpassing the CKB and reaching the QKB. Section~\ref{sec:thermal} illustrates the general results by considering a situation in which the system and controller are initially in thermal equilibrium states, including particular cases of two-level and four-level systems coupled to a thermal spin bath. Finally, conclusions are summarized in Sec.~\ref{sec:conclusions}.

\section{Problem statement and landscape's topological properties}
\label{sec:kinematic analysis}

\begin{figure}
  \includegraphics[width=\columnwidth]{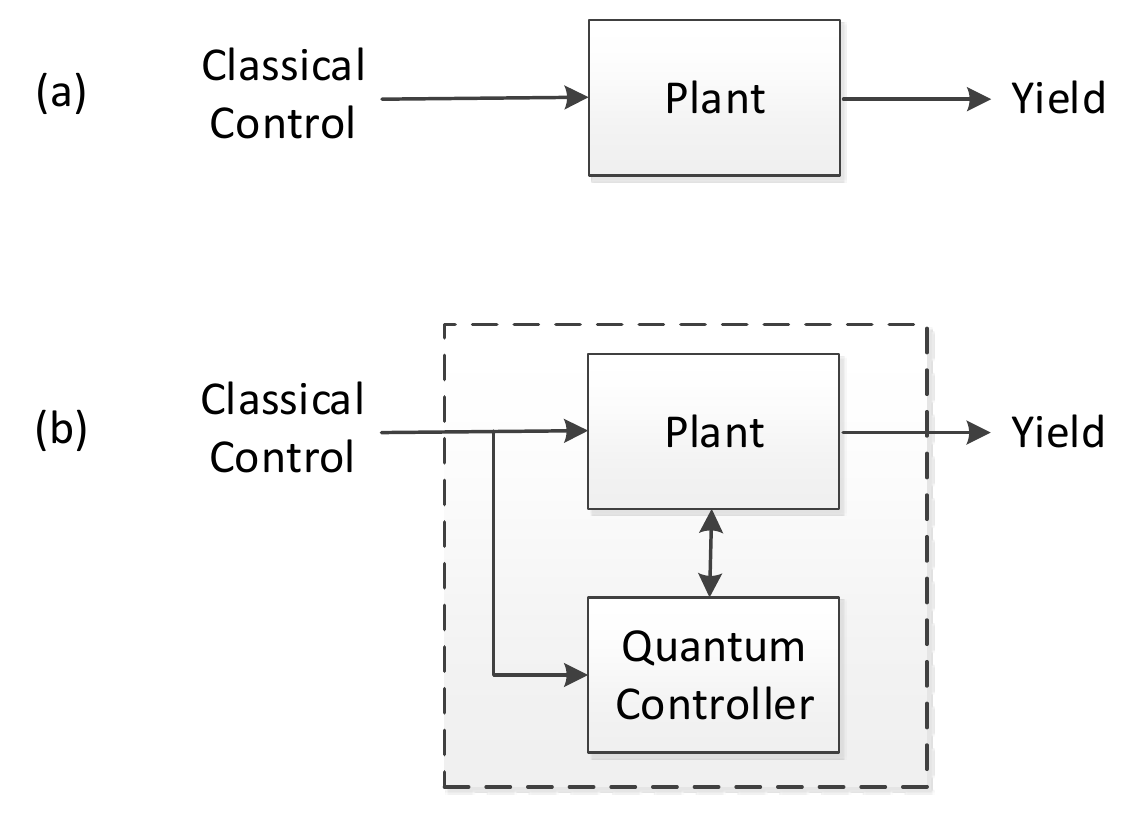}\\
  \caption{Block-diagrams of quantum control systems: (a) a quantum plant manipulated by a classical control; (b) a quantum plant manipulated by a classical control and a quantum controller.}
\label{fig:ControlFrame}
\end{figure}

Generally, control is applied to the quantum plant by interacting it with external agent systems. Depending on whether the agent system is classical or quantum, one can design a classical or quantum controller. As shown in Fig.~\ref{fig:ControlFrame}(a), a classical controller unidirectionally supervises the plant, as usually the backaction from the plant can be ignored. The quantum controller is coherently coupled to the plant, which enables coherent feedback between them, and they can be manipulated by a classical controller, as shown in Fig.~\ref{fig:ControlFrame}(b). Other types of quantum control protocols (e.g., field-mediated coherent feedback~\cite{Wiseman.Milburn.PRA.49.4110.1994, Yanagisawa.Kimura.IEEE-TAC.48.2107.2003, *Yanagisawa.Kimura.IEEE-TAC.48.2121.2003, James.IEEE-TAC.53.1787.2008, Zhang.James.IEEE-TAC.56.1535.2011, Zhang.James.ChinSciBull.57.2200.2012, Zhang.IEEE-TAC.57.1997.2012, Mabuchi.PRA.78.032323.2008, Iida.IEEE-TAC.57.2045.2012, *Crisafulli.OE.21.18371.2013, Kerckhoff.PRL.109.153602.2012, *Kerckhoff.PRX.3.021013.2013} or measurement-based feedback~\cite{Wiseman.Milburn.PRL.70.548.1993, *Wiseman.PRA.49.2133.1994, Lloyd.Viola.PRA.65.010101.2001, Combes.PRA.82.022307.2010, Bushev.PRL.96.043003.2006, *Koch.PRL.105.173003.2010, *Gillett.PRL.104.080503.2010, *Sayrin.Nature.477.73.2011, *Brakhane.PRL.109.173601.2012, *Vijay.Nature.490.77.2012, *Riste.PRL.109.240502.2012, *Campagne-Ibarcq.PRX.3.021008.2013, *Blok.NatPhys.3.189.2014}) can be introduced as well, but they will not be considered in this paper.

Let $N_{\mathrm{s}}$ and $N_{\mathrm{c}}$ be, respectively, the dimensions of the quantum plant and the quantum controller, which together form a composite system of dimension $N = N_{\mathrm{s}}N_{\mathrm{c}}$. The unitary evolution operator (propagator) $U(t) \in \text{U}(N)$ for the composite system satisfies the Schr\"{o}dinger equation:
\begin{equation}
\label{Eq:Sd-equation-enlarged}
    \imath \hbar \dot{U}(t) = \left[ H_{\mathrm{s}}(t) \otimes \mathbb{I}_{\mathrm{c}}
    + \mathbb{I}_{\mathrm{s}} \otimes H_{\mathrm{c}}(t) + H_{\mathrm{int}}(t) \right] U(t) ,
\end{equation}
where $H_{\mathrm{s}}(t)$ and $H_{\mathrm{c}}(t)$ are the Hamiltonians of the system and quantum controller, and $\mathbb{I}_{\mathrm{s}}$ and $\mathbb{I}_{\mathrm{c}}$ are their identity operators, respectively. The interaction Hamiltonian $H_{\mathrm{int}}(t)$ produces an inherent quantum coherent feedback loop between the system and controller~\cite{Lloyd.PRA.62.022108.2000}, through which energy and entropy are exchanged. The Hamiltonians $H_{\mathrm{s}}(t)$, $H_{\mathrm{c}}(t)$, and $H_{\mathrm{int}}(t)$ are tunable through coupling to a time-dependent classical controller (or controllers).

The objective to be optimized is the control yield given by the expectation value of a system observable $\theta$:
\begin{equation}
\label{Eq:U-landscape}
J = \text{Tr} \left[U(t_{\mathrm{f}}) P U^{\dag}(t_{\mathrm{f}}) \Theta \right],
\end{equation}
where $\Theta = \theta \otimes \mathbb{I}_{{\mathrm{c}}}$, $P$ is the initial density matrix of the composite system, and $U(t_{\mathrm{f}})$ is the evolution operator for the composite system at some prescribed final time $t_{\mathrm{f}}$.

For a given observable, the kinematic bounds on the control yield, $J_{\mathrm{max}}$ and $J_{\mathrm{min}}$, are determined by the initial state $P$. We assume that $P$ is a separable state: $P = \rho_{\mathrm{s}} \otimes\rho_{\mathrm{c}}$, where $\rho_{\mathrm{s}}$ and $\rho_{\mathrm{c}}$ are the initial density matrices of the system and controller, respectively. The initial state can, in principle, be a classically correlated state or an entangled state, but these possibilities will not be discussed here.

The evolution operator $U(t_{\mathrm{f}})$ [through its dependence on the Hamiltonians $H_{\mathrm{s}}(t)$, $H_{\mathrm{c}}(t)$, and $H_{\mathrm{int}}(t)$] and, correspondingly, the control yield $J$ [through its dependence on $U(t_{\mathrm{f}})$] are functionals of the control variables. The functional dependence of $J$ on the control variables, through Eq.~\eqref{Eq:U-landscape}, defines the control landscape. If the composite system is evolution-operator controllable, i.e., any $U(t_{\mathrm{f}}) \in \text{U}(N)$ is accessible through Eq.~\eqref{Eq:Sd-equation-enlarged} with some choice of the control variables, then the kinematic bounds $J_{\mathrm{max}}$ and $J_{\mathrm{min}}$ are dynamically reachable. Controllability of the composite system is a reasonable assumption if the quantum controller is of manageable size. As proven in~\cite{Wu.PRA.83.062306.2011, Wu.PRA.86.013405.2012}, if (i) the composite system is evolution-operator controllable, (ii) the Jacobian of the map from the control variables to the evolution operator is full rank, and (iii) the control variables are not constrained, then the analysis of the landscape \eqref{Eq:U-landscape} can be reduced to the so-called kinematic picture, in which the yield $J$ is considered as a function of the evolution operator $U(t_{\mathrm{f}})$ over the unitary group $\text{U}(N)$.

\begin{figure}
  \includegraphics[width=\columnwidth]{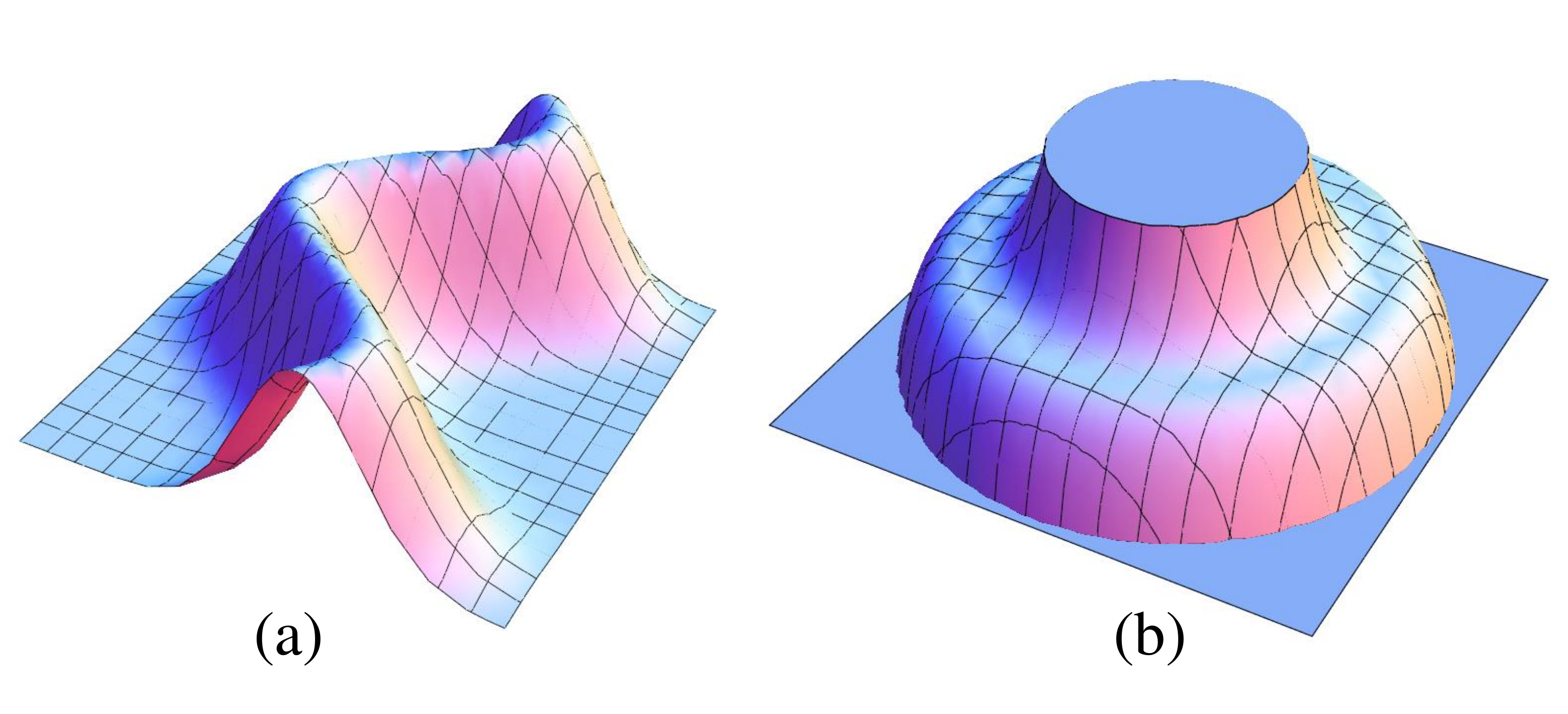}\\
  \caption{(Color online) Schematics of two typical trap-free control landscapes. For both landscapes, the search is in a two-dimensional control space. (a) The maximum submanifold is one-dimensional (with codimension 1) and there are no saddles; (b) the maximum submanifold is two-dimensional (with codimension 0) but there is a one-dimensional saddle submanifold.}
\label{fig:Schematics}
\end{figure}

In previous studies~\cite{Wu.JPA.41.015006.2008, Wu.JMP.49.022108.2008}, the kinematic landscape picture (valid under the above three assumptions) has been used to prove that the critical points of the landscape (\ref{Eq:U-landscape}) can be grouped into a finite number of connected submanifolds of $\text{U}(N)$, each corresponding to a fixed yield value. These critical submanifolds include a unique global maximum submanifold and a unique global minimum submanifold, and the rest are saddle submanifolds (i.e., the landscape has no local optima). Degeneracy structures of $P$ and $\Theta$ determine the number $\mathcal{N}$ of critical submanifolds (or, equivalently, the number $\mathcal{N} - 2$ of saddle submanifolds) and the codimension $D_{\max}$ of the maximum submanifold in the control space. Optimization searches tend to be more efficient on landscapes with smaller $\mathcal{N}$ and $D_{\max}$. For example, comparing the schematic landscapes in Fig.~\ref{fig:Schematics}(a)~and~\ref{fig:Schematics}(b), we see that the former has no saddles, likely leading to faster searches over this landscape in the beginning, but the latter has the maximum submanifold of a higher dimension, which will likely accelerate the search near the top of this landscape as well as likely enable finding optimal solutions that are robust to control noise~\cite{Beltrani.JCP.134.194106.2011, Kosut.PRA.88.052326.2013, Hocker.PRA.90.062309.2014}.

For given degeneracy structures of $P$ and $\Theta$, the landscape topology characteristics $\mathcal{N}$ and $D_{\max}$ can be calculated with the contingency table technique developed in~\cite{Wu.JPA.41.015006.2008, Wu.JMP.49.022108.2008}. For $P$, we consider cases where initial states of the system and controller are pure, mixed, or maximally mixed, which are identified by the spectrum of the density matrix. The density matrix of a pure state has only one nonzero eigenvalue 1, otherwise the state is called mixed. In particular, the state whose density matrix $P = \mathbb{I}/N$ has all identical eigenvalues is called maximally mixed. For simplicity, we consider mixed states whose density matrices have non-degenerate spectra. For $\Theta$, let the observable $\theta$ have $r$ distinct eigenvalues $\bar{\theta}_1 < \cdots < \bar{\theta}_r$ with multiplicities (degeneracy indices) $n_1, \ldots, n_r$, respectively. The corresponding analytical expressions obtained for $\mathcal{N}$ and $D_{\max}$ are listed in Table~\ref{tab:summary} (see~Ref.~\cite{Wu.JMP.49.022108.2008} for derivation details).

It is seen that when the quantum controller is prepared in a mixed state, there are more critical submanifolds on the landscape and the dimension of the maximum critical submanifold is smaller, which typically translates into a higher search effort. Therefore, qualitatively, one should aim to prepare the controller in a pure state. The degeneracy of the target observable's spectrum also affects the $\mathcal{N}$ and $D_{\max}$ values, such that the search efforts tends to be higher when $\theta$ has more distinct eigenvalues or its maximum eigenvalue is less degenerate.

\section{Kinematic bounds on the control yield}
\label{sec:qualitative-analysis}

When a closed quantum system is coupled to a classical controller, the control yield is given by
\begin{equation}
\label{Eq:Classical_landscape}
J^{\mathrm{cl}} = \text{Tr} \left[ U_{\mathrm{s}}(t_{\mathrm{f}}) \rho_{\mathrm{s}}
U_{\mathrm{s}}^{\dag}(t_{\mathrm{f}}) \theta\right] ,
\end{equation}
where $\rho_{\mathrm{s}}$ is the system's initial density matrix, $U_{\mathrm{s}}(t_{\mathrm{f}}) \in \text{U}(N_{\mathrm{s}})$ is the system's evolution operator at the final time $t_{\mathrm{f}}$, and $\theta$ is the target observable. A classical controller can only connect states with the same density matrix spectrum (and, correspondingly, with the same value of entropy)~\cite{Schirmer.JPA.35.4125.2002, *Schirmer.JPA.35.8551.2002, *Albertini.IEEE-TAC.48.1399.2003, Wu.Pechen.Brif.JPA.40.5681.2007}. If the system is initially in a pure state, then a classical controller can evolve it only into other pure states. If the system is initially in a mixed state, then a classical controller will be unable to purify it. These restrictions limit achievable values of $J^{\mathrm{cl}}$. The CKBs, which by definition are the maximum and minimum values of $J^{\mathrm{cl}}$, can be expressed as~\cite{Wu.JPA.41.015006.2008}
\begin{equation}
\label{Eq:CKB}
  J^{\mathrm{cl}}_{\max} = \sum_{k=1}^{N_{\mathrm{s}}} p_k \theta_k, \quad
  J^{\mathrm{cl}}_{\min}=\sum_{k=1}^{N_{\mathrm{s}}} p_{N_{\mathrm{s}}-k+1} \theta_k ,
\end{equation}
where $\{ p_1, \ldots, p_{N_{\mathrm{s}}} \}$ and $\{ \theta_1 , \ldots, \theta_{N_{\mathrm{s}}} \}$ are, respectively, the eigenvalues of $\rho_{\mathrm{s}}$ and $\theta$ in nondecreasing order. Since $\mathrm{Tr} \rho_{\mathrm{s}} = \sum_k p_k = 1$, it is easy to see that the upper CKB $J^{\mathrm{cl}}_{\max}$ is smaller than the upper QKB $\theta_{N_{\mathrm{s}}}$ (and the lower CKB $J^{\mathrm{cl}}_{\min}$ is larger than the lower QKB $\theta_1$) for any \emph{mixed} state $\rho_{\mathrm{s}}$. Pure states are a special case, for which CKBs and QKBs coincide: $J^{\mathrm{cl}}_{\max} = \theta_{N_{\mathrm{s}}}$ and $J^{\mathrm{cl}}_{\min} = \theta_1$ if $\rho_{\mathrm{s}}^2 = \rho_{\mathrm{s}}$.

\begin{table}[t]
\caption{\label{tab:summary}Characteristics of the landscape topology: the number of critical submanifolds, $\mathcal{N}$, and the codimension of the maximum submanifold, $D_{\max}$. The expressions for $\mathcal{N}$ and $D_{\max}$ are obtained for different types of initial states of the system and controller: pure, non-degenerate mixed, and maximally mixed (MM); ``---'' indicates that no analytical formula was derived for this case.}
\begin{ruledtabular}
\begin{tabular}{cc|cc}
$\rho_{\mathrm{s}}$ & $\rho_{\mathrm{c}}$ & $\mathcal{N}$ &~$D_{\max}$~\\
\hline
Pure & Pure & $~r~$ & $2(N_{\mathrm{s}}-n_1) N_{{\mathrm{c}}}$ \\
Mixed & Pure & ${\displaystyle r^{N_{\mathrm{s}}} }$ & $2(N_{\mathrm{s}}-n_1) N$ \\
MM & Pure & ${\displaystyle \frac{(N_{\mathrm{s}}+r-1)!}{N_{\mathrm{s}}!(r-1)!} }$ &
$2(N_{\mathrm{s}}-n_1) N$ \\
\hline
Pure & Mixed & ${\displaystyle r^{N_{{\mathrm{c}}}} }$ & $2(N_{\mathrm{s}}-n_1) N_{{\mathrm{c}}}^2$ \\
Mixed & Mixed & ${\displaystyle \frac{N!}{(N_{{\mathrm{c}}} n_1)! \cdots
(N_{{\mathrm{c}}} n_{r})!} }$ & $\left( N_{\mathrm{s}}^2 - \sum_i n_i^2 \right) N_{{\mathrm{c}}}^2$ \\
MM & Mixed & --- & $\left( N_{\mathrm{s}}^2 - \sum_i n_i^2 \right) N_{{\mathrm{c}}}^2$\\
\hline
Pure & MM & ${\displaystyle \frac{(N_{\mathrm{c}}+r-1)!}{N_{\mathrm{s}}!(r-1)!}}$ &
$2(N_{\mathrm{s}}-n_1) N_{{\mathrm{c}}}^2$ \\
Mixed & MM & --- & $\left( N_{\mathrm{s}}^2 - \sum_i n_i^2 \right) N_{{\mathrm{c}}}^2$ \\
\end{tabular}
\end{ruledtabular}
\end{table}

The only way to improve the control yield for mixed initial states beyond the CKB is to transfer entropy to some other system, for example, through coupling to a quantum controller (another possibility, which we do not consider in this paper, is the use of measurements~\cite{Vilela-Mendes.PRA.67.053404.2003, *Roa.PRA.73.012322.2006, *Zhang.PRA.73.032101.2006, *Pechen.PRA.74.052102.2006, *Shuang.JCP.126.134303.2007, *Dong.JCP.129.154103.2008, *Jacobs.NJP.12.043005, *Konrad.Uys.PRA.85.012102.2012}, including the application of measurement-based feedback control \cite{Wiseman.Milburn.PRL.70.548.1993, *Wiseman.PRA.49.2133.1994, Lloyd.Viola.PRA.65.010101.2001, Combes.PRA.82.022307.2010, Bushev.PRL.96.043003.2006, *Koch.PRL.105.173003.2010, *Gillett.PRL.104.080503.2010, *Sayrin.Nature.477.73.2011, *Brakhane.PRL.109.173601.2012, *Vijay.Nature.490.77.2012, *Riste.PRL.109.240502.2012, *Campagne-Ibarcq.PRX.3.021008.2013, *Blok.NatPhys.3.189.2014}). For the general situation involving coupling to a quantum controller, the control landscape is given by Eq.~\eqref{Eq:U-landscape}, while the landscape of Eq.~\eqref{Eq:Classical_landscape} can be considered as a special case when $N_{\mathrm{c}}=1$. Let $\{ P_1, \ldots, P_{N} \}$ and $\{ \Theta_1 , \ldots, \Theta_{N} \}$ be, respectively, the eigenvalues of $P$ and $\Theta$ in nondecreasing order. Note that the composite system undergoes a unitary evolution, which preserves the spectrum of its state. Previous studies have shown~\cite{Wu.JPA.41.015006.2008, Wu.JMP.49.022108.2008} that critical values of the control yield $J$ of Eq.~\eqref{Eq:U-landscape} are given by
\begin{equation}
\label{Eq:J-critical}
J_{\sigma} = \sum_{k=1}^{N} P_{\sigma(k)}  \Theta_k , \quad
\sigma \in \mathfrak{S}_{N} .
\end{equation}
Here, $\mathfrak{S}_{N}$ is the group of all permutations on $N$ indices, i.e., the symmetric group. In particular, the maximum and minimum values of $J$ are
\begin{equation}
\label{Eq:JmaxJmin}
J_{\max} = \sum_{k=1}^{N} P_{k}  \Theta_k, \quad
J_{\min} = \sum_{k=1}^{N} P_{N-k+1} \Theta_k ,
\end{equation}
corresponding to the identity and reverse-order permutations, respectively. Since $\mathrm{Tr} P = \sum_k P_k = 1$, the value of $J_{\max}$ ($J_{\min}$) is always bounded from above (below) by the maximum (minimum) eigenvalue of $\Theta$, which is the upper (lower) QKB. Note that the upper QKB is $\Theta_{N} = \theta_{N_{\mathrm{s}}} = \bar{\theta}_r$ and the lower QKB is $\Theta_1 = \theta_1 = \bar{\theta}_1$. Next, we will derive the condition for the general kinematic bounds of Eq.~\eqref{Eq:JmaxJmin} to surpass the CKBs of Eq.~\eqref{Eq:CKB} as well as that to reach the QKBs.

\subsection{Surpassing the CKB}

Let $J_{\sigma_0}$ be a critical value of the control yield, which corresponds to the permutation $\sigma_0$ that arranges the eigenvalues of $P$ in the following order:
\begin{equation}
\label{Eq:order-sigma-0}
p_1 q_1, \ldots, p_1 q_{N_{\mathrm{c}}}; \ldots ;
p_{N_{\mathrm{s}}} q_1, \ldots, p_{N_{\mathrm{s}}} q_{N_{\mathrm{c}}} ,
\end{equation}
where $\{ p_1, \ldots, p_{N_{\mathrm{s}}} \}$ and $\{ q_1 , \ldots, q_{N_{\mathrm{c}}} \}$ are, respectively, the eigenvalues of $\rho_{\mathrm{s}}$ and $\rho_{\mathrm{c}}$ in nondecreasing order. Substituting the order of Eq.~\eqref{Eq:order-sigma-0} into Eq.~\eqref{Eq:J-critical}, we obtain:
\begin{eqnarray}
\label{Eq:J-sigma-0}
J_{\sigma_0} & = & p_1 \left( q_1 \Theta_1 + \cdots
+ q_{N_{\mathrm{c}}} \Theta_{N_{\mathrm{c}}} \right) + \cdots \nonumber \\
&& + p_{N_{\mathrm{s}}} \left( q_1 \Theta_{N - N_{\mathrm{c}} + 1}
+ \cdots + q_{N_{\mathrm{c}}} \Theta_{N} \right) \nonumber \\
& = & \left( p_1 \theta_1 + \cdots + p_{N_{\mathrm{s}}} \theta_{N_{\mathrm{s}}} \right)
\left( q_1 + \cdots + q_{N_{\mathrm{c}}} \right) ,
\end{eqnarray}
where we used the fact that each eigenvalue of $\theta$ enters $N_{\mathrm{c}}$ times into the spectrum of $\Theta$. Since $\mathrm{Tr} \rho_{\mathrm{c}} = \sum_k q_k = 1$, we immediately see that the critical value of Eq.~\eqref{Eq:J-sigma-0} is equal to the upper CKB of Eq.~\eqref{Eq:CKB}: $J_{\sigma_0} = J^{\mathrm{cl}}_{\max}$.

It is useful to arrange the terms in $J_{\sigma_0}$ according to the degeneracy of $\theta$. Specifically, we divide the spectrum of $P$ into bands $\mathcal{B}_i$ ($i = 1, \ldots, r)$ such that all elements of $\mathcal{B}_i$ multiply the same distinct eigenvalue $\bar{\theta}_i$ in $J_{\sigma_0}$:
\begin{equation}
\label{Eq:J-sigma-0-bands}
J_{\sigma_0} = \sum_{i = 1}^r \left( \sum_{P_k \in \mathcal{B}_i} P_k \right) \bar{\theta}_i .
\end{equation}
Based on the order of Eq.~\eqref{Eq:order-sigma-0}, it is easy to see that the bands $\mathcal{B}_i$ are
\begin{eqnarray*}
\mathcal{B}_1 & = & \{ p_1 q_1, \ldots, p_1 q_{N_{\mathrm{c}}}; \ldots;
p_{\mu_1} q_1, \ldots, p_{\mu_1} q_{N_{\mathrm{c}}} \}, \\
   & \vdots &  \\
\mathcal{B}_r & = & \{p_{\mu_{r-1}+1} q_1, \ldots, p_{\mu_{r-1}+1} q_{N_{\mathrm{c}}}; \ldots;
p_{\mu_r} q_1, \ldots, p_{\mu_r} q_{N_{\mathrm{c}}} \} ,
\end{eqnarray*}
where
\begin{equation}
\label{Eq:band_division}
\mu_i = \sum_{l = 1}^i n_l , \quad i = 1, \ldots, r,
\end{equation}
i.e., $\mu_i$ is the combined multiplicity of distinct eigenvalues $\{ \bar{\theta}_1, \ldots , \bar{\theta}_i \}$ or, equivalently, the total number of eigenvalues of $\theta$, which are equal to or less than $\bar{\theta}_i$ (e.g., $\mu_r = N_{\mathrm{s}}$).

\begin{figure}[t]
  \includegraphics[width=0.9\columnwidth]{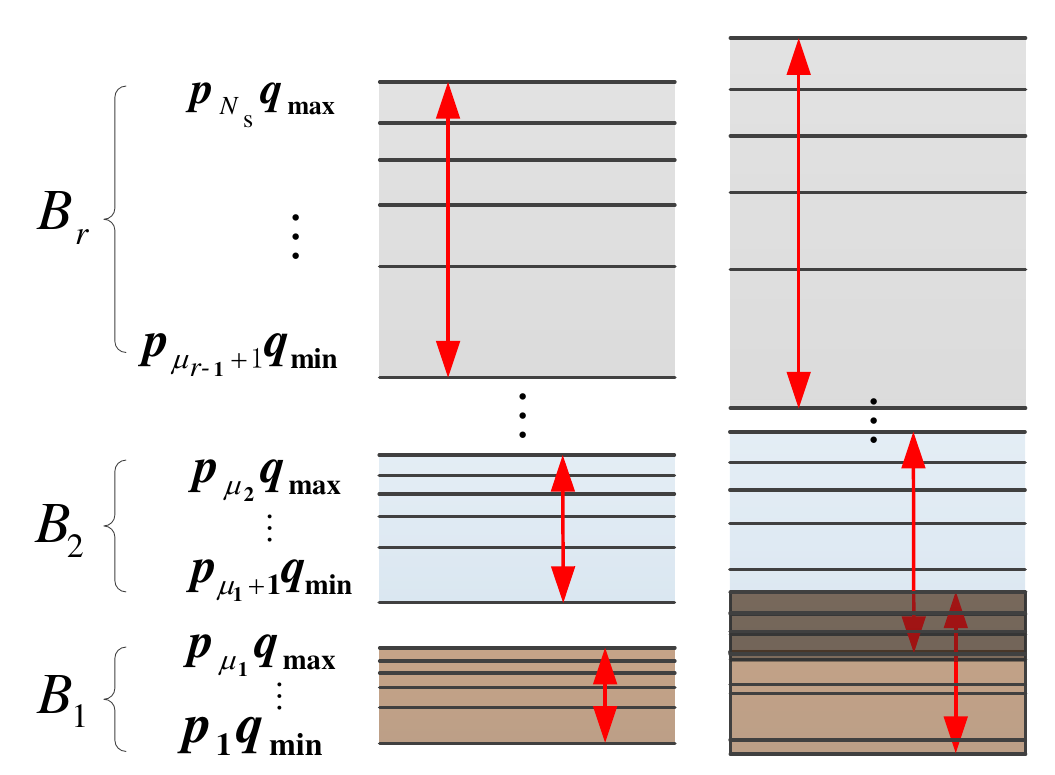}\\
  \caption{(Color online) A schematic diagram of typical spectra of $P$, illustrating condition~\eqref{condition-1} for surpassing the upper CKB on the control yield. The spectrum of $P$ is decomposed into bands $\mathcal{B}_i$ ($i = 1, \ldots, r)$ such that all elements of $\mathcal{B}_i$ multiply the same distinct eigenvalue $\bar{\theta}_i$ in $J_{\sigma_0}$ [cf.~Eq.~\eqref{Eq:J-sigma-0-bands}]. When none of the bands overlap (left), the upper kinematic bound $J_{\max}$ remains the same as the upper CKB $J_{\sigma_0} = J^{\mathrm{cl}}_{\max}$. Only when at least two of the bands overlap (right), $J_{\max}$ exceeds the upper CKB.}
\label{fig:overlap}
\end{figure}

Since the value of $J_{\sigma_0}$ is invariant to the order of eigenvalues of $P$ in each band $\mathcal{B}_i$, it is always possible to assume that all $P_k \in \mathcal{B}_i$ are arranged in nondecreasing order for each $\mathcal{B}_i$. Correspondingly, if none of the bands $\mathcal{B}_i$ overlap (i.e., if $p_{\mu_i} q_{N_{\mathrm{c}}} \leq p_{\mu_i + 1} q_1$ $\forall i$), then \emph{all} eigenvalues of $P$ in $J_{\sigma_0}$ are arranged in nondecreasing order, which means that $J_{\sigma_0} = J_{\max}$ and, consequently, the CKB is not broken: $J_{\max} = J_{\sigma_0} = J^{\mathrm{cl}}_{\max}$. However, when at least two of these bands overlap with each other, $\sigma_0$ does not arrange the eigenvalues of $P$ in nondecreasing order, and hence $J_{\sigma_0}$ is not the maximum of the control yield~\eqref{Eq:U-landscape}. In such a case, $J_{\sigma_0} = J^{\mathrm{cl}}_{\max}$ is smaller than $J_{\max}$ of Eq.~\eqref{Eq:JmaxJmin}, which means that the upper CKB can be surpassed. Thus, we can conclude that the upper CKB can be surpassed if and only if
\begin{equation}
\label{condition-1}
p_{\mu_i} q_{\max} > p_{\mu_i + 1} q_{\min}
\end{equation}
for at least one $i \in \{1, 2, \ldots, r-1\}$, where we denoted $q_{\min} = q_1$ and $q_{\max} = q_{N_{\mathrm{c}}}$. Figure~\ref{fig:overlap} shows a schematic diagram of the spectrum of $P$ for the cases when none of the bands overlap and two of the bands overlap.

Condition~\eqref{condition-1} depends on the spectra of $\rho_{\mathrm{s}}$ and $\rho_{\mathrm{c}}$, as well as on the degeneracy structure of $\theta$. To better understand these dependencies, we define the spectral bandwidth of the controller's state $\rho_{\mathrm{c}}$ as
\begin{equation}
\label{Eq:Control_Bandwidth}
  B_{\mathrm{c}} = \ln\frac{q_{\max}}{q_{\min}} .
\end{equation}
In particular, if $q_{\min}=0$, then $B_{\mathrm{c}} \rightarrow \infty$. Similarly, we define the band gaps in the spectrum of $\rho_{\mathrm{s}}$ as
\begin{equation}
\label{Eq:System_Bandgaps}
  g_k = \ln\frac{p_{k+1}}{p_k},\quad k=1,2,\cdots,N_{\mathrm{s}}-1.
\end{equation}
For example, if the spectrum of $\rho_{\mathrm{s}}$ has a degeneracy such that $p_k = p_{k+1}$, then $g_k = 0$. Using these definitions, condition~\eqref{condition-1} for surpassing the upper CKB can be reformulated as the following theorem:
\begin{theorem}
\label{TH-1}
The upper CKB can be surpassed if and only if the spectral bandwidth $B_{\mathrm{c}}$ of the controller's state is greater than the minimum band gap among $\{g_{\mu_1},\ldots,g_{\mu_{r-1}}\}$, i.e.,
\begin{equation}
\label{Eq:Condition}
  B_{\mathrm{c}} > \min_{1 \leq i < r} g_{\mu_i} .
\end{equation}
\end{theorem}

The same conclusion can be extended to the analysis of the lower kinematic bound by arranging the eigenvalues of $P$ in nonincreasing order. Let us denote
\begin{equation}
\label{Eq:band_division_min}
\nu_i = N_{\mathrm{s}} - \mu_i = \sum_{l = i+1}^r n_l, \quad i = 1, \ldots, r-1 ,
\end{equation}
i.e., $\nu_i$ is is the combined multiplicity of distinct eigenvalues $\{ \bar{\theta}_{i+1}, \ldots , \bar{\theta}_r \}$ or, equivalently, the total number of eigenvalues of $\theta$, which are greater than $\bar{\theta}_i$. Then the lower CKB can be surpassed if and only if
\begin{equation}
\label{condition-1_min}
p_{\nu_i} q_{\max} > p_{\nu_i + 1} q_{\min}
\end{equation}
for at least one $i \in \{1, 2, \ldots, r-1\}$ or, equivalently,
\begin{equation}
\label{Eq:Condition_min}
  B_{\mathrm{c}} > \min_{1 \leq i < r} g_{\nu_i}.
\end{equation}
If the spectrum of $\theta$ is non-degenerate, then both conditions~\eqref{Eq:Condition} and \eqref{Eq:Condition_min} reduce to
\begin{equation}
\label{Eq:Condition_nd}
  B_{\mathrm{c}} > \min_{1 \leq k < N_{\mathrm{s}}} g_k.
\end{equation}

From Theorem~\ref{TH-1}, it is easy to derive the following two corollaries:
\begin{corollary}
\label{CR-1}
If at least one $p_{\mu_i}$ ($p_{\nu_i}$) is nonzero for $1 \leq i < r$, the upper (lower) CKB can always be surpassed when $q_{\min} = 0$, i.e., when $\rho_{\mathrm{c}}$ has at least one zero eigenvalue.
\end{corollary}
{\bf Proof:} When $q_{\min} = 0$, condition~\eqref{condition-1} reads $p_{\mu_i} q_{\max} > 0$, which is  satisfied if at least one $p_{\mu_i}$ is nonzero. The proof for the lower kinematic bound is analogous, using condition~\eqref{condition-1_min}.

\begin{corollary}
\label{CR-2}
The CKB cannot be surpassed if $\rho_{\mathrm{s}}$ is a pure state or $\rho_{\mathrm{c}}$ is maximally mixed.
\end{corollary}
{\bf Proof:} If $\rho_{\mathrm{s}}$ is pure, then $p_{N_{\mathrm{s}}} = 1$ and $p_k = 0$ $\forall k \neq N_{\mathrm{s}}$. Consequently, $p_{\mu_i} = 0$ and $p_{\nu_i} = 0$ $\forall i \in \{1, 2, \ldots, r-1\}$, and therefore conditions~\eqref{condition-1} and \eqref{condition-1_min} can never be satisfied. If $\rho_{\mathrm{c}}$ is maximally mixed, then $q_{\max} = q_{\min} = 1/N_{\mathrm{c}}$, and the controller's bandwidth is zero, which can never be greater than any band gap of $\rho_{\mathrm{s}}$.

The results in Corollary~\ref{CR-2} have simple physical interpretations. When the system is in a pure state, the CKB and QKB coincide and thus no improvement beyond CKB is possible. When the state of the controller is maximally mixed, its entropy is maximal, and hence no flow of entropy from the system to the controller is possible, regardless of the system's state. More generally, Theorem~\ref{TH-1} indicates that, to improve the control yield, the controller's spectral bandwidth $B_{\mathrm{c}}$ should be sufficiently large. This means that $\rho_{\mathrm{c}}$ should be a low entropy state. Note that in control of closed quantum systems with classical fields, the control pulse bandwidth is the key resource that determines the minimum time necessary to perform the optimal dynamics and saturate the CKB \cite{Lloyd.Montangero.PRL.113.010502.2014}, but does not affect the value of the bound. In comparison, the spectral bandwidth of the quantum controller's state is a key resource that impacts how much the value of the kinematic bound on the control yield can be increased beyond the CKB.

\subsection{Reaching the QKB}

Defined as the maximum and minimum eigenvalues of the target observable, the QKBs represent the ultimate physical limit on the achievable yield. Consider the maximum value of $J$ given by Eq.~\eqref{Eq:JmaxJmin} (the analysis for the minimum value of $J$ is completely analogous). We decompose the summation in the general expression~\eqref{Eq:JmaxJmin} for $J_{\max}$ into two parts:
\begin{equation}
\label{Eq:Jmax_double_sum}
J_{\max} = \sum_{k=1}^{N - n_r N_{\mathrm{c}}} P_k \Theta_k
+ \left( \sum_{k = N - n_r N_{\mathrm{c}} + 1}^{N} P_k \right) \bar{\theta}_r ,
\end{equation}
where all $\Theta_k$'s in the first summation are smaller than the maximum eigenvalue $\Theta_{N} = \theta_{N_{\mathrm{s}}} = \bar{\theta}_r$. Evidently, $J_{\max}$ can exactly reach the upper QKB $\bar{\theta}_r$ if and only if the first term in Eq.~\eqref{Eq:Jmax_double_sum} is zero and the sum over $P_k$'s in the second term is one, which requires that all nonzero $P_k$'s are in the second term. Since the eigenvalues $\{P_k\}$ are in nondecreasing order, this implies that the QKB is attainable if and only if the number of nonzero eigenvalues of $P$ is no greater than the multiplicity $n_r N_{\mathrm{c}}$ of $\bar{\theta}_r$ in the spectrum of $\Theta$, as summarized in the following theorem.

\begin{theorem}
\label{TH-2}
The upper QKB can be exactly reached: $J_{\max} = \bar{\theta}_r$, if and only if
\begin{equation}
\label{condition-2}
(N_{\mathrm{s}} - N_{\mathrm{s} 0}) (N_{\mathrm{c}} - N_{\mathrm{c} 0}) \leq n_r N_{\mathrm{c}},
\end{equation}
where $N_{\mathrm{s} 0}$ and $N_{\mathrm{c} 0}$ are, respectively, the nullities of $\rho_{\mathrm{s}}$ and $\rho_{\mathrm{c}}$. The same condition, with $n_r$ replaced by $n_1$ in~\eqref{condition-2}, can be obtained for $J_{\min}$ to reach the lower QKB $\bar{\theta}_1$.
\end{theorem}
Note that the sum of the nullity (the number of zero eigenvalues) of a matrix and its rank is equal to the number of its columns, so the right-hand side of Eq.~\eqref{condition-2} can be expressed as $(N_{\mathrm{s}} - N_{\mathrm{s} 0}) (N_{\mathrm{c}} - N_{\mathrm{c} 0}) = \mathrm{rank}(\rho_{\mathrm{s}})\, \mathrm{rank}(\rho_{\mathrm{c}})$.

In distinction to Theorem~\ref{TH-1}, condition~\eqref{condition-2} depends on $\rho_{\mathrm{s}}$ and $\rho_{\mathrm{c}}$ only through the integral indices $N_{\mathrm{s}0}$ and $N_{\mathrm{c}0}$, but not through the magnitude of the associated eigenvalues. The dependence on the target observable $\theta$ is again through the degeneracy of its spectrum, but Eq.~\eqref{condition-2} depends on only one multiplicity $n_r$ (or $n_1$). In particular, it would be more difficult to satisfy the condition of Theorem~\ref{TH-2} for an observable whose maximum (minimum) eigenvalue is not degenerate. An important example of such a situation is the problem of achieving the maximum degree of purification of the system's state, in which case $\theta$ is a projector onto a pure state (with only one nonzero eigenvalue 1). For this special case, Theorem~\ref{TH-2}, with $n_r = 1$ in Eq.~\eqref{condition-2}, states the condition for achieving complete purification (in this sense, the purification problem is no different from the control yield maximization for any other observable with non-degenerate maximum eigenvalue). Thus, Theorem~\ref{TH-2} is a generalization of a previously obtained condition on the reachability of maximum purification~\cite{Ticozzi.Viola.SciRep.4.5192.2014}.

It is also interesting to consider a situation when the system consists of $m_{\mathrm{s}}$ qubits ($N_{\mathrm{s}} = 2^{m_{\mathrm{s}}}$), the controller consists of $m_{\mathrm{c}}$ qubits ($N_{\mathrm{c}} = 2^{m_{\mathrm{c}}}$), and the invariant subspace corresponding to the maximum eigenvalue of $\theta$ consists of $m_r$ qubits ($n_r = 2^{m_r}$). An example of such an observable is a projector on the subspace of $m_r$ qubits ($m_r < m_{\mathrm{s}}$). In this situation, condition~\eqref{condition-2} can be expressed as
\begin{equation}
\label{condition-2-qubits}
m_r + m_{\mathrm{c}} \geq S_0(\rho_{\mathrm{s}}) + S_0(\rho_{\mathrm{c}}) ,
\end{equation}
where $S_0(\rho) = \log_2 \mathrm{rank}(\rho)$ is the quantum Hartley entropy~\cite{Bengtsson.Zyczkowski.2006.book}, which is a measure of the information lost due to the mixedness of the state. Using the form~\eqref{condition-2-qubits} of condition~\eqref{condition-2}, Theorem~\ref{TH-2} can be interpreted in information-theoretic terms: the QKB can be exactly reached if and only if the sum of the number of qubits, $m_r + m_{\mathrm{c}}$, is not less than the number of bits of information lost due to the mixedness of the system's and controller's states.

A general conclusion of Theorem~\ref{TH-2} is that, in order to reach the QKB, the system's and/or controller's initial state needs to have sufficiently many zero eigenvalues. In the already discussed special case of a pure $\rho_{\mathrm{s}}$, the CKB and QKB coincide, which means that the QKB is always reachable. Indeed, when $\rho_{\mathrm{s}}$ is a pure state, $N_{\mathrm{s}} - N_{\mathrm{s} 0} = 1$ and condition~\eqref{condition-2} is always satisfied. A more interesting special case is when $\rho_{\mathrm{c}}$ is a pure state (i.e., $N_{\mathrm{c}} - N_{\mathrm{c} 0} = 1$). From Theorem~\ref{TH-2}, it is easy to derive the following two corollaries:
\begin{corollary}
\label{CR-3}
The QKB on the control yield is reachable if the controller is initially in a pure state and its dimension satisfies $N_{\mathrm{c}} \geq (N_{\mathrm{s}} - N_{\mathrm{s} 0} ) / n_{\mathrm{r}}$.
\end{corollary}
\begin{corollary}
\label{CR-4}
For any initial state of the system and any target observable, the QKB on the control yield is reachable if the controller is initially in a pure state and its dimension satisfies $N_{\mathrm{c}} \geq N_{\mathrm{s}}$.
\end{corollary}
It is interesting to compare Corollary~\ref{CR-4} to a previously obtained result on Kraus-map controllability of a quantum system coupled to a quantum controller~\cite{Wu.Pechen.Brif.JPA.40.5681.2007}. Assuming that the composite system is evolution-operator controllable (which is also required for the kinematic bounds obtained in this paper to be dynamically reachable), the result in~\cite{Wu.Pechen.Brif.JPA.40.5681.2007} states that the system is Kraus-map controllable if the quantum controller is initially in a pure state and its dimension satisfies $N_{\mathrm{c}} \geq N_{\mathrm{s}}^2$. Thus, the condition of Corollary~\ref{CR-4} for reaching the QKB on the control yield is weaker (i.e., a smaller controller is needed) than that for Kraus-map controllability. In other words, the QKB for observable control can be reached without all Kraus maps of the system being accessible.

In practice, it often happens that all levels of the system and controller are populated, but many of the populations are so small that they can be treated nearly as zero. In such a case, while the QKB cannot be perfectly reached, it can still be approached very closely.

\section{Kinematic bounds for thermal systems}
\label{sec:thermal}

Realistic physical systems often evolve from a thermal equilibrium state. Let the system and controller be initially in thermal states with temperatures $T_{\mathrm{s}}$ and $T_{\mathrm{c}}$, respectively:
\begin{equation}
\label{Eq:Thermal_state}
\rho_{\mathrm{s}} = Z^{-1}_{\mathrm{s}} e^{-H_{\mathrm{s}} / k_{\mathrm{B}} T_{\mathrm{s}} } , \quad
\rho_{\mathrm{c}} = Z^{-1}_{\mathrm{c}} e^{-H_{\mathrm{c}} / k_{\mathrm{B}} T_{\mathrm{c}} } .
\end{equation}
Here, $H_{\mathrm{s}}$ and $H_{\mathrm{c}}$ are the respective Hamiltonians at the initial time, $Z_{\mathrm{s}} = \mathrm{Tr}\left(e^{-H_{\mathrm{s}} / k_{\mathrm{B}} T_{\mathrm{s}} }\right)$ and $Z_{\mathrm{c}} = \mathrm{Tr}\left(e^{-H_{\mathrm{c}} / k_{\mathrm{B}} T_{\mathrm{c}} } \right)$ are the respective partition functions, and $k_{\mathrm{B}}$ is the Boltzmann constant. The states~\eqref{Eq:Thermal_state} are diagonal in the respective energy eigenbases, and energy level populations are density matrix eigenvalues. In this section, we study how the kinematic bounds on the control yield depend on the system's and controller's energy spectra and temperatures.

Let $\{ \hbar\omega_1, \ldots, \hbar\omega_{N_{\mathrm{s}}} \}$ be eigenvalues of $H_{\mathrm{s}}$ in nonincreasing order; correspondingly, the eigenvalues of $\rho_{\mathrm{s}}$ are $\{ p_1, \ldots, p_{N_{\mathrm{s}}} \}$ in nondecreasing order, where $p_k = Z^{-1}_{\mathrm{s}} e^{-\hbar\omega_k / k_{\mathrm{B}} T_{\mathrm{s}} }$ ($k = 1,\ldots,N_{\mathrm{s}}$). Let $\hbar \Omega_{\min}$ and $\hbar \Omega_{\max}$ be, respectively, the minimum and maximum eigenvalues of $H_{\mathrm{c}}$; correspondingly, the maximum and minimum eigenvalues of $\rho_{\mathrm{c}}$ are $q_{\max} = Z^{-1}_{\mathrm{c}} e^{-\hbar \Omega_{\min} / k_{\mathrm{B}} T_{\mathrm{c}} }$ and $q_{\min} = Z^{-1}_{\mathrm{c}} e^{-\hbar \Omega_{\max} / k_{\mathrm{B}} T_{\mathrm{c}} }$, respectively. Substituting these expressions for $p_k$, $q_{\max}$, and $q_{\min}$ into Eq.~\eqref{condition-1}, we find that for thermal states the upper CKB can be surpassed if and only if
\begin{equation}
\label{condition-1-thermal-1}
(\Omega_{\max} - \Omega_{\min}) / T_{\mathrm{c}} >
(\omega_{\mu_i} - \omega_{\mu_i + 1}) / T_{\mathrm{s}} ,
\end{equation}
for at least one $i \in \{1, 2, \ldots, r-1\}$, where $\{\mu_i \}$ are defined by Eq.~\eqref{Eq:band_division} according to the degeneracy structure of the $\theta$ spectrum. We also obtain thermal-state expressions for the spectral bandwidth of the controller's state:
\begin{equation}
\label{Eq:Control_Bandwidth_thermal}
  B_{\mathrm{c}} = \frac{ \hbar (\Omega_{\max}-\Omega_{\min})}{k_{\mathrm{B}} T_{\mathrm{c}}}
\end{equation}
and the band gaps in the spectrum of $\rho_{\mathrm{s}}$:
\begin{equation}
\label{Eq:System_Bandgaps_thermal}
  g_k = \frac{\hbar(\omega_k - \omega_{k+1})}{k_{\mathrm{B}} T_{\mathrm{s}}}, \quad k=1,2,\cdots,N_{\mathrm{s}}-1.
\end{equation}
Using the form~\eqref{condition-1-thermal-1} of condition~\eqref{condition-1}, we formulate another corollary of Theorem~\ref{TH-1}.
\begin{corollary}
\label{CR-5}
When the system and controller are initially in thermal states with temperatures $T_{\mathrm{s}}$ and $T_{\mathrm{c}}$, respectively, the upper CKB can be surpassed if and only if
\begin{equation}
\label{Eq:Thermal_CKB}
\Omega_{\max} - \Omega_{\min} > \frac{T_{\mathrm{c}}}{T_{\mathrm{s}}}
\min_{1 \leq i < r} (\omega_{\mu_i} - \omega_{\mu_i + 1}) .
\end{equation}
\end{corollary}

We see that, for a thermal system and controller at given temperatures, the bandwidth of the controller's energy spectrum, $\Omega_{\max} - \Omega_{\min}$, has to be sufficiently large to break the CKB. On the other hand, it is easier to break the CKB when the system's energy spectrum is dense. Also, according to Corollary~\ref{CR-5}, the hotter is the controller, the more difficult it becomes to surpass the CKB. This is not surprising since entropy of a thermal controller increases with the temperature. Ultimately, in the limit $T_{\mathrm{c}} \rightarrow \infty$, the controller is in the maximally mixed state and the CKB cannot be surpassed in agreement with Corollary~\ref{CR-2}. In the opposite limit of zero $T_{\mathrm{c}}$, the controller is in the ground state and the CKB can always be surpassed (for non-zero $T_{\mathrm{s}}$) in agreement with Corollary~\ref{CR-1}. It is also worth noting that, for a finite-temperature system, only a zero-temperature controller can exactly reach the QKB, provided that $N_{\mathrm{c}} \geq N_{\mathrm{s}} / n_r$. Regarding the effect of $T_{\mathrm{s}}$, the hotter is the system, the tighter is the CKB and hence the less demanding is the condition for surpassing it. Below, we illustrate these results by considering two specific examples.

\subsection{A two-level system coupled to a thermal spin bath}
\label{sec:example-1}

Consider a two-level quantum system coupled to a controller which is a collection of $M$ identical spins. Let the angular frequency of the system be $\omega_{\mathrm{s}}$ and that of the identical spins in the controller be $\omega_{\mathrm{c}}$. The respective Hamiltonians are
\begin{equation}
H_{\mathrm{s}} = \frac{\hbar\omega_{\mathrm{s}}}{2} \sigma_z, \quad
H_{\mathrm{c}} = \frac{\hbar\omega_{\mathrm{c}}}{2} \sum_{\ell=1}^M \sigma_z^{(\ell)},
\end{equation}
where $\sigma_z$ is the Pauli matrix and $\sigma_z^{(\ell)} = \mathbb{I}_2^{\otimes (\ell-1)} \otimes \sigma_z\otimes \mathbb{I}_2^{\otimes (M-\ell)}$. The controller's Hamiltonian $H_{\mathrm{c}}$ has $M+1$ equally spaced eigenvalues:
\begin{equation}
E_k = k \hbar\omega_{\mathrm{c}}, \quad k = -\frac{M}{2},\ldots,\frac{M}{2},
\end{equation}
with multiplicities $\text{C}_M^0,\text{C}_M^1,\ldots,\text{C}_M^M$, respectively. The target observable is chosen to be $\theta = \sigma_z$ (the same results would also be obtained for $\theta = \sigma_x$ or $\theta = \sigma_y$). One can verify by symmetry that $J_{\min} = -J_{\max}$, so it is sufficient to investigate only the upper kinematic bound $J_{\max}$.

Let the system and controller be initially in thermal states of Eq.~\eqref{Eq:Thermal_state} with temperatures $T_{\mathrm{s}}$ and $T_{\mathrm{c}}$, respectively. The spectra of $\rho_{\mathrm{s}}$ and $\rho_{\mathrm{c}}$ depend, respectively, on dimensionless parameters
\begin{equation}\label{Eq:Parameters}
\lambda_{\mathrm{s}} = \frac{\hbar\omega_{\mathrm{s}}}{k_{\mathrm{B}} T_{\mathrm{s}}}, \quad
\lambda_{\mathrm{c}} = \frac{\hbar\omega_{\mathrm{c}}}{k_{\mathrm{B}} T_{\mathrm{c}}}.
\end{equation}
According to Corollary~\ref{CR-5}, the CKB can be surpassed if and only if
\begin{equation}
\label{Eq:2-level-CKB-condition}
  M \lambda_{\mathrm{c}} > \lambda_{\mathrm{s}}.
\end{equation}

\begin{figure}[t]
  \includegraphics[width=\columnwidth]{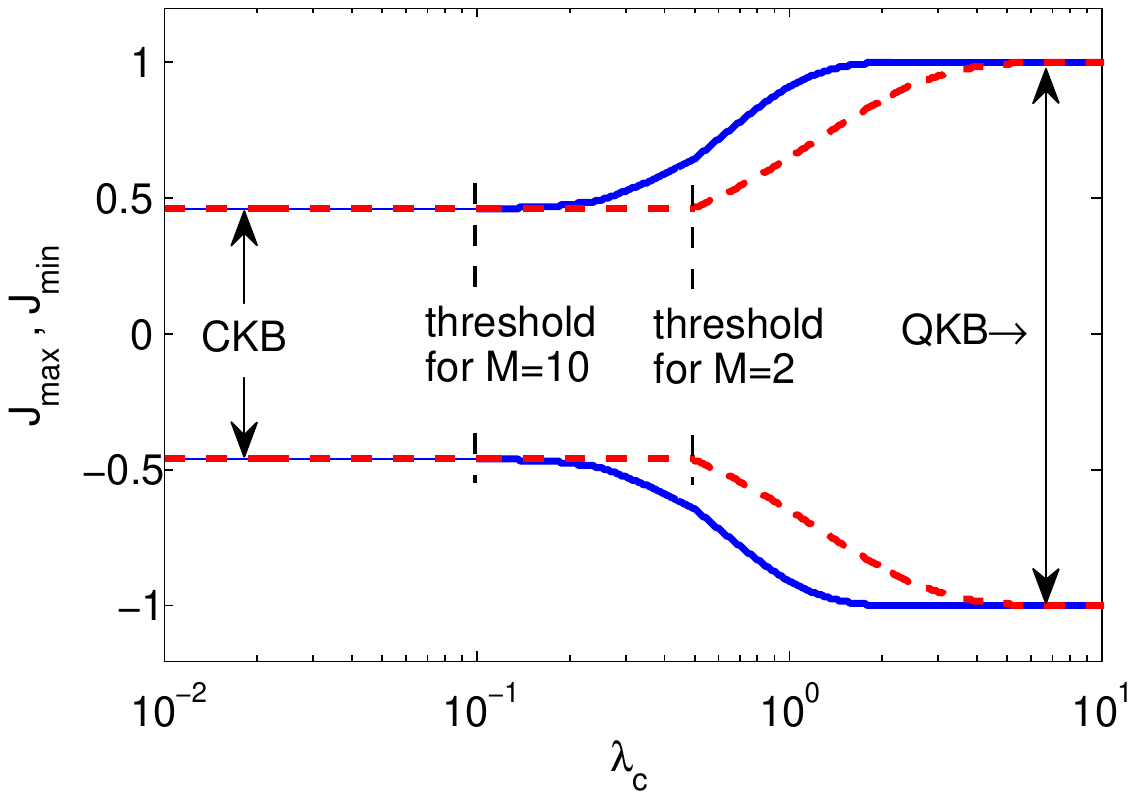}\\
  \caption{(Color online) The kinematic bounds on the control yield $J$ for a two-level quantum system coupled to a collection of $M$ identical spins. Both the system and controller are initially in thermal equilibrium states. The system's target observable is $\sigma_z$. The bounds $J_{\max}$ and $J_{\min}$ are shown as functions of the controller's parameter $\lambda_{\mathrm{c}}$ for a fixed $\lambda_{\mathrm{s}} = 1$ and two values of $M$. The blue solid curve corresponds to $M=10$ and the red dashed curve to $M=2$. The CKB $J^{\mathrm{cl}}_{\max} = \tanh(\lambda_{\mathrm{s}}/2) \approx 0.4621$ is surpassed for $\lambda_{\mathrm{c}} > \lambda_{\mathrm{s}}/M$.}
\label{fig:KB}
\end{figure}

Figure~\ref{fig:KB} shows the dependence of the kinematic bounds $J_{\max}$ and $J_{\min}$ on the controller's parameter $\lambda_{\mathrm{c}}$ for a fixed $\lambda_{\mathrm{s}} = 1$ and two values of $M$ ($M = 2$ and $M = 10$). As long as $\lambda_{\mathrm{c}} \leq \lambda_{\mathrm{s}}/M$, the CKB holds: $J_{\max} = J^{\mathrm{cl}}_{\max} = \tanh(\lambda_{\mathrm{s}}/2) \approx 0.4621$. When $\lambda_{\mathrm{c}}$ exceeds the threshold value $\lambda_{\mathrm{s}}/M$, the bound $J_{\max}$ surpasses the CKB and starts to grow. Since thermal states have no zero eigenvalues (for non-zero temperature), condition~\eqref{condition-2} of Theorem \ref{TH-2} cannot be exactly satisfied for any finite value of $\lambda_{\mathrm{c}}$. Nevertheless, for $\lambda_{\mathrm{c}} \gg 1$, populations of the higher energy levels are nearly zero, and the kinematic bound asymptotically approaches the QKB: $J_{\max} \rightarrow \theta_2 = 1$. Comparing the curves for different numbers of the controller's spins ($M = 2$ versus $M = 10$), we see that both the threshold value for surpassing the CKB, $\lambda_{\mathrm{c}} = \lambda_{\mathrm{s}}/M$, and a value of $\lambda_{\mathrm{c}}$ at which $J_{\max}$ can be considered practically equal to the QKB (e.g., 99\% yield) are lower for larger $M$. This implies that a larger controller is more effective in expanding the kinematic bounds. However, a caveat is that finding proper controls with a larger quantum controller is more difficult than with a smaller one.

\subsection{A two-spin system coupled to a thermal spin bath}
\label{sec:example-2}

Consider a quantum system of two identical spins, whose Hamiltonian is
\begin{equation}
\label{Eq:Ham-2-spins}
H_{\mathrm{s}} = \frac{\hbar\omega_{\mathrm{s}}}{2} (\sigma_z\otimes \mathbb{I}_2+ \mathbb{I}_2\otimes \sigma_z)
=\left(
   \begin{array}{cccc}
   \hbar\omega_{\mathrm{s}}   &  &  &  \\
      & 0 &  &  \\
      &  & 0 &  \\
      &  &  & -\hbar\omega_{\mathrm{s}} \\
   \end{array}
 \right)
\end{equation}
in the basis $\{ |\uparrow \uparrow \rangle, |\uparrow \downarrow \rangle, |\downarrow \uparrow \rangle, |\downarrow \downarrow \rangle \}$. The system is coupled to the same controller (a collection of $M$ identical spins) as described in Sec.~\ref{sec:example-1}. The system and controller are initially in thermal states of Eq.~\eqref{Eq:Thermal_state} with temperatures $T_{\mathrm{s}}$ and $T_{\mathrm{c}}$, respectively. We consider two target observables $\theta = \Pi_0$ and $\theta = \Pi_1$, where
\begin{equation}
\label{Eq:theta-projectors}
\Pi_0=\left(
   \begin{array}{cccc}
   0   &  &  &  \\
      & 1 &  &  \\
      &  & 1 &  \\
      &  &  & 0 \\
   \end{array}
 \right)
\quad \text{and} \quad
\Pi_1=\left(
   \begin{array}{cccc}
   1  &  &  &  \\
      & 0 &  &  \\
      &  & 0 &  \\
      &  &  & 0 \\
   \end{array}
 \right)
\end{equation}
are the projectors on the subspace $\{ |\uparrow \downarrow \rangle, |\downarrow \uparrow \rangle \}$ and the state $|\uparrow \uparrow \rangle$, respectively. For these observables, we investigate the upper kinematic bound. According to Eq.~\eqref{Eq:band_division}, we obtain $\mu_1 = 2$, $\mu_2 = 4$ for $\theta = \Pi_0$ and $\mu_1 = 3$, $\mu_2 = 4$ for $\theta = \Pi_1$.

Since for both choices of $\theta$ we have $r = 2$, condition~\eqref{Eq:Condition} takes the form:
\begin{equation}
\label{Eq:2-spin-CKB-condition}
 B_{\mathrm{c}} > g_{\mu_1} .
\end{equation}
Using the notation introduced in Eq.~\eqref{Eq:Parameters}, the controller's bandwidth [cf.~Eq.~\eqref{Eq:Control_Bandwidth_thermal}] is $B_{\mathrm{c}} = M \lambda_{\mathrm{c}}$, and the band gaps in the spectrum of $\rho_{\mathrm{s}}$ [cf.~Eq.~\eqref{Eq:System_Bandgaps_thermal}] are $g_1 = \lambda_{\mathrm{s}}$, $g_2 = 0$, $g_3 = \lambda_{\mathrm{s}}$. For $\theta = \Pi_0$, we have $g_{\mu_1} = g_2 = 0$, and hence the upper CKB can be surpassed if and only if
\begin{equation}
\label{Eq:2-spin-CKB-condition-P0}
 M \lambda_{\mathrm{c}} > 0,
\end{equation}
which is always true for any finite bath temperature. On the other hand, for $\theta = \Pi_1$, we have $g_{\mu_1} = g_3 = \lambda_{\mathrm{s}}$, and hence the upper CKB can be surpassed if and only if
\begin{equation}
\label{Eq:2-spin-CKB-condition-P1}
 M \lambda_{\mathrm{c}} > \lambda_{\mathrm{s}},
\end{equation}
which requires that the bandwidth of the controller's energy spectrum is sufficiently large.

\begin{figure}
  \includegraphics[width=\columnwidth]{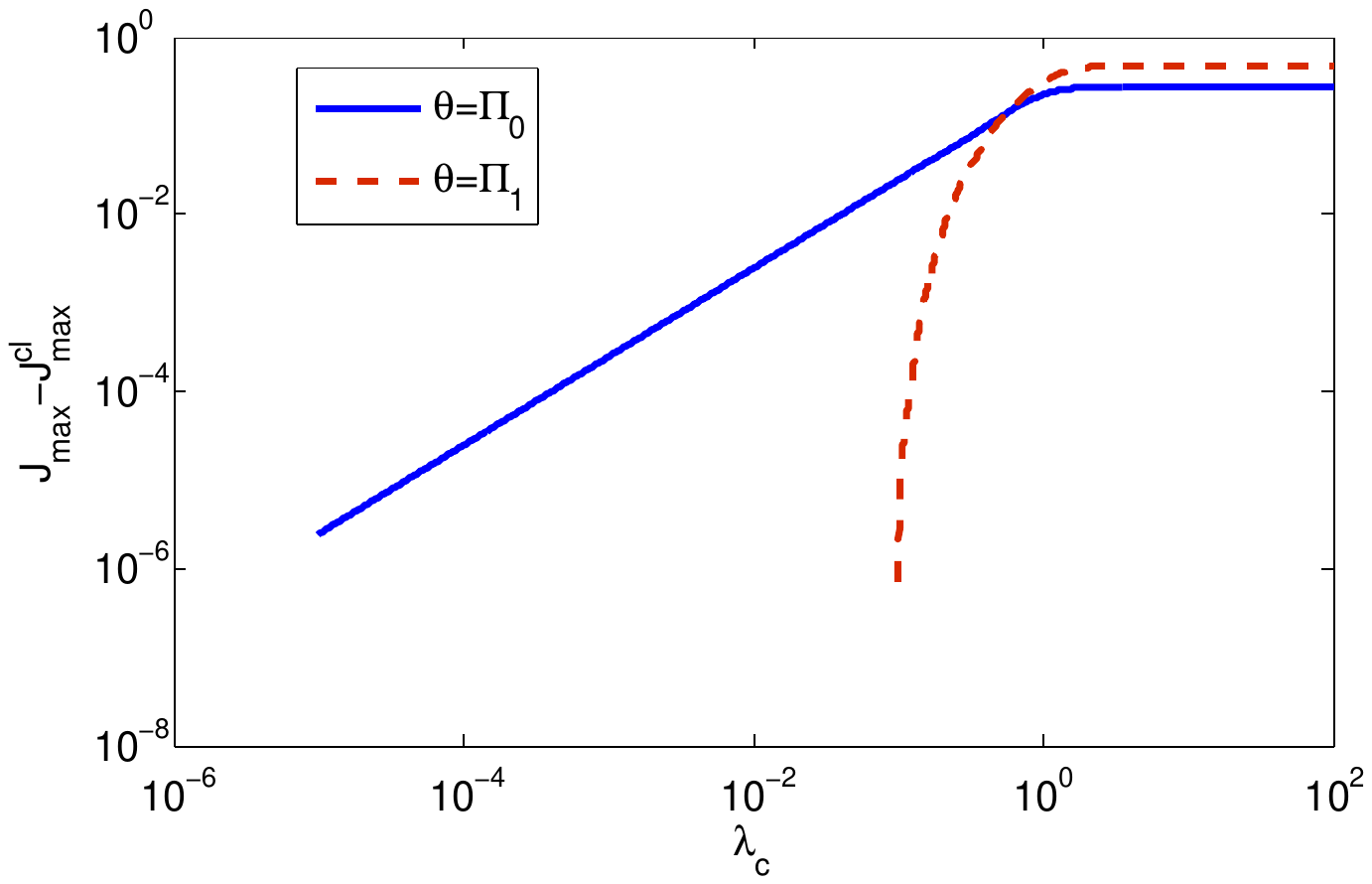}\\
  \caption{(Color online) The difference $J_{\max} - J_{\max}^{\mathrm{cl}}$ as a function of the controller's parameter $\lambda_{\mathrm{c}}$, for a two-spin quantum system (with a fixed $\lambda_{\mathrm{s}} = 1$) coupled to a collection of $M = 10$ identical spins. Both the system and controller are initially in thermal equilibrium states. The blue solid curve corresponds to the system's target observable $\theta$ chosen as the projector $\Pi_0$ on the subspace $\{ |\uparrow \downarrow \rangle, |\downarrow \uparrow \rangle \}$, and the green dashed curve corresponds to the system's target observable $\theta$ chosen as the projector $\Pi_1$ on the state $|\uparrow \uparrow \rangle$. For $\theta = \Pi_0$, the kinematic bound $J_{\max}$ is always above the CKB, while for $\theta = \Pi_1$, $J_{\max}$ only surpasses the CKB when $\lambda_{\mathrm{c}}$ exceeds the threshold value $\lambda_{\mathrm{s}}/M = 0.1$.}
\label{fig:KB2}
\end{figure}

Figure~\ref{fig:KB2} shows the difference between $J_{\max}$ and the upper CKB $J_{\max}^{\mathrm{cl}}$ as a function of $\lambda_{\mathrm{c}}$ (with a fixed $\lambda_{\mathrm{s}} = 1$ and $M = 10$) for the two choices of $\theta$. It is clearly seen that, in accordance with the above analysis, $J_{\max}$ surpasses the CKB only above the threshold value $\lambda_{\mathrm{c}} = \lambda_{\mathrm{s}}/M = 0.1$ for $\theta = \Pi_1$, but $J_{\max}$ is always above the CKB for $\theta = \Pi_0$. This distinction illustrates the crucial role played by the degeneracy structure of the $\theta$ spectrum. For $\theta = \Pi_0$, the amount by which $J_{\max}$ exceeds the CKB decreases as $\lambda_{\mathrm{c}}$ goes to zero (corresponding to narrowing of the controller's spectral bandwidth). For both objectives, the QKB can be closely approached when $\lambda_{\mathrm{c}}$ is very large.

\section{Conclusions}
\label{sec:conclusions}

We investigated kinematic bounds on the control yield, defined as the expectation value of a system observable, for a general situation where the system of interest (the quantum plant) is coupled to both an external classical field (a classical controller) and an auxiliary quantum system (a quantum controller). When a closed quantum system is coupled only to a classical controller, its dynamics is limited to unitary orbits, and the control yield is limited by the CKB that depends on the initial state. The main question explored in this work is to what degree the use of a quantum controller can aid in expanding the yield bounds. We answered this question by deriving two main results: the necessary and sufficient condition for the bound to surpass the CKB, as well as that for the bound to exactly reach the QKB which constitutes the ultimate physical limit.

The condition for surpassing the CKB is expressed in terms of eigenvalues of the system's and controller's initial density matrices and also depends on the degeneracy structure of the target observable's spectrum. In order to satisfy this condition, it is desirable to prepare the controller in a state with a large spectral bandwidth (which translates into the energy spectral bandwidth for a thermal state). In comparison, the spectral bandwidth of a classical control field is a key factor determining the minimum time necessary to saturate the CKB, but it cannot affect the value of the bound.

The condition for reaching the QKB does not depend on the magnitude of eigenvalues of the initial density matrices, but rather is expressed in terms of their ranks as well as the multiplicity of the maximum (minimum) eigenvalue of the target observable. For the system and controller consisting of qubits, this condition has an information-theoretic interpretation, i.e., the number of controller qubits should be not less than the number of bits of information lost due to the mixedness of the system's and controller's states (quantified by the sum of their quantum Hartley entropies). If the controller is initially in a pure state, the QKB is reachable for any initial system state and any target observable, provided that the dimension of the controller is not less than that of the system. This condition on the controller's dimension is much milder than that for Kraus-map controllability, i.e., reachability of the ultimate physical limit for observable control does not require the capability to generate all Kraus maps of the system.

The general results as well as examples of thermal initial states indicate that a larger quantum controller is favorable for expanding the kinematic bounds on the control yield, as it can more effectively absorb entropy from the system. However, dynamic reachability of the kinematic bounds is guaranteed only if the composite system is evolution-operator controllable, which may have practical consequences. Despite this practical limitation, our findings are likely to provide a basis for a better understanding of control yields observed in experiments as well as help to establish the principles for designing effective quantum controllers.

%

\acknowledgments
The authors thank Dr.~Mohan Sarovar for useful discussions. RBW acknowledges support from NSFC Grant No. 60904034, 61374091 and 61134008. CB acknowledges support from the Laboratory Directed Research and Development program at Sandia National Laboratories. Sandia National Laboratories is a multi-program laboratory managed and operated by Sandia Corporation, a wholly owned subsidiary of Lockheed Martin Corporation, for the U.S. Department of Energy's National Nuclear Security Administration under contract DE-AC04-94AL85000. MRJ acknowledges support from the Australian Research Council Centre of Excellence for Quantum Computation and Communication Technology (project number CE110001027), and AFOSR Grant FA2386-12-1-4075. HR acknowledges partial support from NSF Grant No. CHE-1058644 and ARO-MURI Grant No. W911NF-11-1-2068.

\bibliographystyle{apsrev4-1}
%

\end{document}